\documentclass[conference]{IEEEtran}
\IEEEoverridecommandlockouts
\usepackage{cite}
\usepackage{amsmath,amssymb,amsfonts}
\usepackage{algorithmic}
\usepackage{graphicx}
\usepackage{textcomp}
\usepackage{xcolor}
\def\BibTeX{{\rm B\kern-.05em{\sc i\kern-.025em b}\kern-.08em
		T\kern-.1667em\lower.7ex\hbox{E}\kern-.125emX}}

\usepackage{url}
\usepackage{units} %nicefrac
\graphicspath{{./Figures/}}
\DeclareGraphicsExtensions{.pdf,.jpeg,.png}
\usepackage{color, soul}

\begin{document}
	
	\title{DER Information Unaware Coordination via Day-ahead Dynamic Power Bounds\\
		\thanks{The work presented in this paper was partially supported under ARPA-E award DE-AR0000697. Copyright Notice: 978-1-7281-6127-3/20/\$31.00 ©2020 IEEE}
	}
	% Use this to place sponsorships
	%\thanksto{*The work presented in this paper was partially supported under ARPA-E award DE-AR0000697.}
	
	%\and
	%\and
	%\and
	%\and
	%\and
	
	\author{\IEEEauthorblockN{Thomas Navidi\IEEEauthorrefmark{1},
			Chloe Leblanc\IEEEauthorrefmark{1},
			Abbas El Gamal\IEEEauthorrefmark{1} and 
			Ram Rajagopal\IEEEauthorrefmark{1}\IEEEauthorrefmark{2}
			% and Author n.5\IEEEauthorrefmark{4}
		}
		\IEEEauthorblockA{\IEEEauthorrefmark{1} Electrical Engineering, Stanford University}
		\IEEEauthorblockA{\IEEEauthorrefmark{2} Civil and Environmental Engineering, Stanford University\\
			% Name of the organization B,
			\{tnavidi, chloel, abbas, ramr\}@stanford.edu}
		% abbas@ee.stanford.edu 
		% and 
		% ramr@stanford.edu} 	
	}
	
	\maketitle
	
	\begin{abstract}
		
		%Traditionally, reliability and voltage quality in distribution grids are ensured via transformer power ratings and voltage management assets. Under this paradigm, future grids with deep flexible loads, storage, and solar penetrations would require costly equipment upgrades. These upgrades can be mitigated via coordination of loads and storage. In earlier work, a 2-layer DER coordination architecture was shown to achieve close to optimal performance despite significantly delayed communication to a central coordinator. Inspired by this finding, we propose a day-ahead coordination planning tool that uses customer requested power profile ranges to generate \emph{day-ahead dynamic power rating bounds} at each transformer. This scheme differs from our earlier work in that: (i) the DSO does not impose or know consumer objectives, (ii) we consider both flexible loads and storage, and (iii) to provide scalability, we use a data-driven linearized network model. Simulations using the IEEE 123-bus network show that with 60\% solar, 60\% residential EVs and 10\% storage penetrations, the uncoordinated approach incurs rating violations at 37 of 86 transformers and 10X higher squared voltage deviation, while our approach incurs only 12 rating violations and maintains nearly the same voltage deviation as without DERs.
		
		Reliability and voltage quality in distribution networks have been achieved via a combination of transformer power rating satisfaction and voltage management asset control. To maintain reliable operation under this paradigm, however, future grids with deep DER penetrations would require costly equipment upgrades. These upgrades can be mitigated via judicious coordination of DER operation. Earlier work has assumed a hierarchical control architecture in which a global controller (GC) uses detailed power injection and DER data and knowledge of DER owners' objectives to determine setpoints that local controllers should follow in order to achieve reliable and cost effective grid operation. Having such detailed data and assuming knowledge of DER owners' objectives, however, are often not desirable or possible. In an earlier work, a 2-layer DER coordination architecture was shown to achieve close to optimal performance despite infrequent (e.g., once per day) communication to a global controller. Motivated by this work, this paper proposes a day-ahead coordination scheme that uses forecasted power profile ranges to generate day-ahead dynamic power rating bounds at each transformer. Novel features of this scheme include: (i) the GC knows only past node power injection data and does not impose or know DER owner objectives, (ii) we use bounds that ensure reliable operation to guide the local controllers rather than setpoint tracking, and (iii) we consider electric vehicle (EV) charging in addition to storage. Simulations using the IEEE 123-bus network show that with random placements of 50\% solar, 50\% EVs and only 10\% storage penetrations, the uncoordinated approach incurs rating violations at nearly all 86 transformers and results in 10 times higher voltage deviation, while our approach incurs only 12 rating violations and maintains almost the same voltage deviations as before the addition of solar and EVs.
		
		% Accuracy of linear model is within %1. 
		% Perfect foresight controller can reduce violations to 0 with high PV penetration and distributed storage
		
	\end{abstract}
	

		\vspace{-3mm}
	\section{Introduction}
	\vspace{-1mm}
	
%	A grand challenge of future electric grids is how to best coordinate  (DERs), , to allow renewables to replace fossil fuel sources without unduly increasing costs. 

	Maintaining reliable and high quality electricity supply to consumers on a distribution grid is traditionally achieved via satisfaction of transformer power ratings and the use of voltage management assets. The introduction of distributed energy resources (DERs), such as rooftop photovoltaics (PV), energy storage, and flexible loads such as electric vehicles (EVs), however, disrupts this paradigm. Distributed PV generates excess power during peak daylight hours causing back-feeding, which increases voltage levels. Many existing assets are unidirectional, thus, not designed to work under back-feeding and can further exacerbate the voltage rise rather than help regulate it~\cite{Sandia_PV_regulators, PV_regulators_journal}. Even bidirectional voltage regulators suffer from high solar variability causing rapid switching, which reduces their life expectancy.
	%Grids with high solar penetration such as Hawaii's have already had to remove incentives such as net-metering to mitigate this problem~\cite{hawaii}, and the growth of rooftop PV is expected to increase rapidly throughout the US~\cite{DOE_REPORT} meaning the problem is sure to spread. 
	Moreover, both residential and commercial EV charging can cause significant power spikes~\cite{EV_stats_times}, resulting in excessive transformer aging~\cite{EV_impacts, EV_transformerAge}. %This problem will be exacerbated in the future due to the expected increase of EV adoption~\cite{DOE_REPORT}.
		
	Addressing the above challenges under the current operation paradigm will lead to costly asset upgrades. %An NREL study~\cite{PV_impacts} describes methods to reduce the cost of asset upgrades to manage distributed PV with a focus on power factor correction using solar inverters, but finds that costly upgrades may be necessary.
	In~\cite{EVCharging_Costs}, the breakdown of costs when installing new EV charging infrastructure shows that upgrading grid assets can be very costly. A report from the CPAU~\cite{PaloAltoUtility-Report} specifically mentions both voltage fluctuations from distributed PV and transformer overloading due to EV charging as concerns that will require costly infrastructure upgrades. %In order for renewable energy to become economically competitive with fossil fuel sources, these costly equipment upgrades must be kept to a minimum.
	
	%Previous work on coordinating DERs for grid reliability can fall into three categories: centralized control \cite{DMPC}, distributed control \cite{boyd_messaging, junjie, consensus} and 
	%2-layer control \cite{Bi-level_Bernstein, PacketizedEnergy}. %The main disadvantages of fully centralized or distributed controllers involve the difficulty establishing frequent communication of large amounts of data to a central controller or communicating in peer to peer networks, which do not currently exist in distribution grids. Furthermore, issues surrounding device ownership and data privacy create conflicts when operating under these schemes. A more comprehensive discussion on the advantages of a 2-layer architecture is presented in \cite{CDCp}. However, not all 2-layer architectures are set up to avoid the challenges that a single architecture faces, particularly, the difficulty in establishing frequent communication between controllers. 
	
	The adverse effects of high DER penetration can be mitigated by judicious coordination of storage and flexible loads in the distribution network. Recent work on coordination of DERs typically involves a hierarchical control architecture with data exchange occurring between a global controller (GC) and local controllers (LCs) at the DERs, e.g., see~\cite{Bi-level_Bernstein, setpoint_track_Bernstein, PacketizedEnergy, kyle, CDCp}. The work in~\cite{Bi-level_Bernstein, PacketizedEnergy} assumes frequent and low delay communication between the global and local controllers, which may not be practical using smart meters and consumer broadband networks. The work in~\cite{kyle, CDCp}, however, has shown that such frequent communication may not be needed, demonstrating close to optimal performance even with once per day communication between the GC and LCs due to the ability of each LC to accurately forecast its own demand and act on it in real time. Most recent work on DER coordination assumes that the GC knows past power injection data, storage SOC, and charging rates, as well as the objectives of all LCs in the system, and uses this knowledge to determine optimal power injection setpoints for each local controller to follow to achieve close to optimal power injections. This includes data-driven approaches such as~\cite{data_ML_local} in which the LCs learn to mimic an offline centralized controller, and~\cite{data_opf_guo} in which a centralized stochastic OPF dictates the operation of its resources using forecast error distributions. Such detailed knowledge of DER data and LC objectives, however, is not desirable or even possible as DER owners may wish to keep their data and objectives private. Moreover, most of the aforementioned recent work on DER coordination does not consider EV charging. 
			
	In this paper, we propose a day-ahead DER coordination scheme that significantly reduces the need for data exchange between the control layers and does not require any knowledge of DER owners' objectives or detailed DER data, e.g., SOC of storage or EV charging rate. Each day, the GC uses past power injection data from smart meters to compute target upper and lower power injections for the next day instead of setpoints as in previous work. It, then minimizes a weighted sum of the squared bus voltage deviations and distance from the power injection targets subject to power flow constraints to determine the upper and lower bounds, which are sent to the LCs. Upon receiving the bounds, each LC minimizes a combination of its own objective and deviation from the bounds subject to the DER constraints. We explore only real power control in this paper because we assume reactive power control is handled automatically by the inverter according to IEEE std. 1547~\cite{IEEE_standard_1547} as is done in practice. Our scheme considers both flexible loads in the form of electric vehicles (EVs) in addition to storage.

	Through simulation with real residential data, we compare the results using our coordination scheme to the setpoint tracking scheme in~\cite{CDCp} and two benchmarks: (i) DERs respond intelligently to time of use (TOU) prices without knowledge of the bounds, and (ii) DERs respond to both prices and the static transformer power rating bounds. We demonstrate that our dynamic bounds enable the coordination of DERs to promote distribution grid reliability, reducing the need for asset upgrades over the benchmarks with similar voltage and transformer rating violations to the tracking scheme in~\cite{CDCp}, which require much more DER information and knowledge of DER owners' objectives.
	
	%The rest of the paper is organized as follows. The next section introduces the models and assumptions used throughout the paper including the data-driven linearized network model. Section~\ref{Architecture} provides a detailed description of the network power bounds scheduler, along with its potential interaction with markets, and a description of the local controllers. Section~\ref{Results} describes our implementation and simulation results on the IEEE 123-bus distribution network~\cite{IEEE_feeders}.
	
	\vspace{-2mm}
	\section{Models and assumptions} \label{models}
\begin{figure}[htpb]
		\centering
		\includegraphics[width=2.5in]{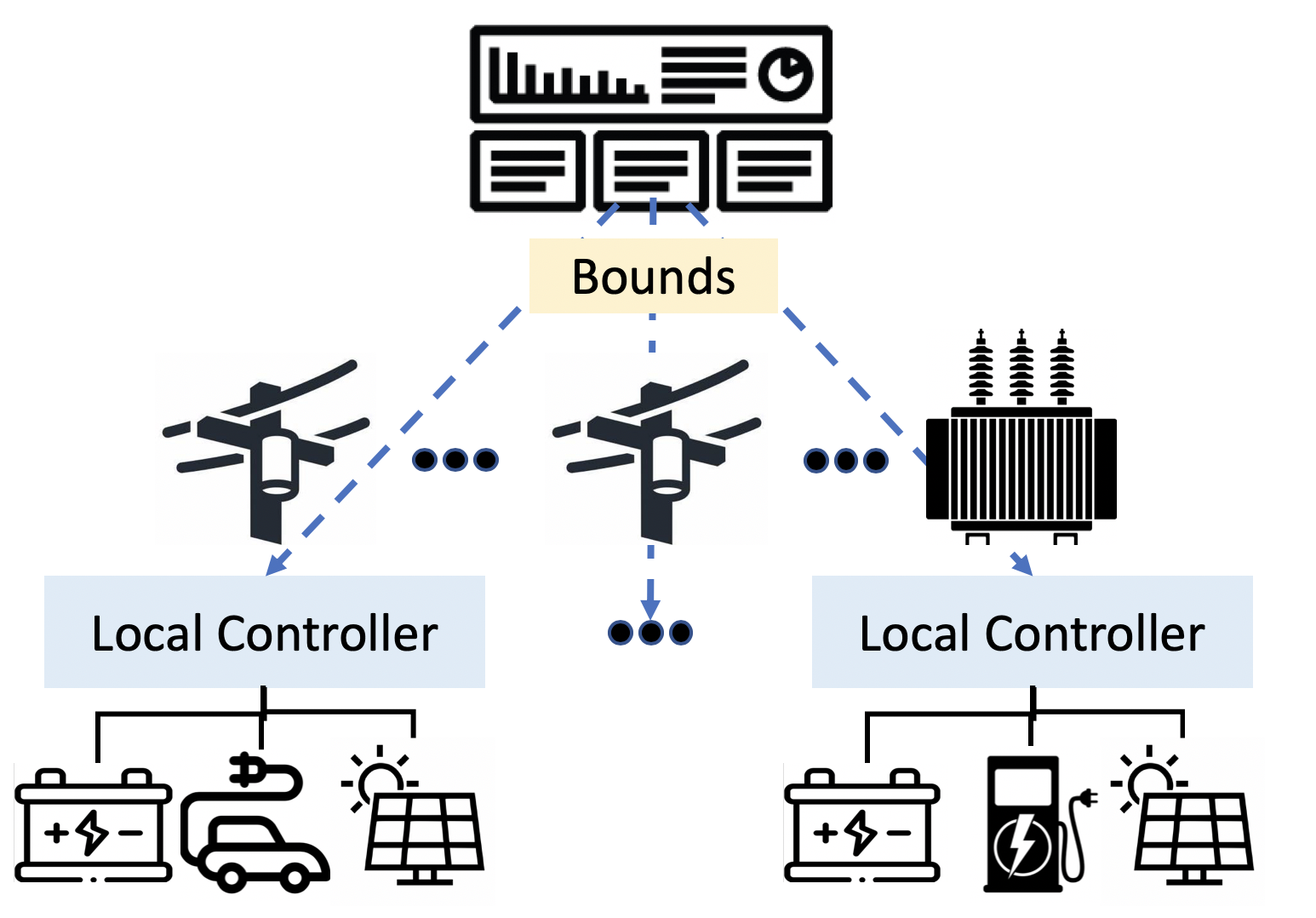}
		\vspace{-2mm}
		\caption{Illustration of the architecture with the global controller that computes the bounds, local controllers, residential DERs, and larger commercial DERs with fast chargers.}
		\label{arch_fig}
		\vspace{-4mm}
	\end{figure}

	\vspace{-1mm}
	We consider a distribution grid with randomly distributed solar generation and EVs. The solar and storage penetrations are defined as percentages of the total network demand, and solar is randomly distributed as in~\cite{CDCp}. %The variables used in the optimization algorithms are given in Table~\ref{varTable}.
	The general architecture we propose is shown in Figure~\ref{arch_fig}. 
	
		%%%architecture figure shows EV and storage
	\vspace{-4mm}
\medskip	
	
	\noindent{\bf EV charging model:} %We sample random distributions to generate charging profiles for the EVs in the grid. 
	We use Gaussian mixture models with parameters determined from~\cite{EV_stats_sim} to sample the percentage of cars charging each day, the starting time, initial charge, and final charge for each EV in the network. The total number of EVs in the network is calculated from the number of houses in the network times the EV penetration percent. We set aside the 4 largest nodes in the network to have fast charging stations with a total capacity of 120 chargers. EVs with start times before 10 AM have a high probability of charging at a fast charging station while cars starting at other times are more likely to charge at home or a parking lot. The probabilities are tuned so that approximately 15\% of EVs are fast charging each day. Fast chargers have a maximum charging power of 40 kW while residential chargers can charge up to 6 kW. We assume the charge rate can be set anywhere from 0 to the maximum, and the amount of time available for charging is 50\% longer than the minimum required time in order to provide some degree of controllability to the LCs for scheduling EV charging.
	
	\noindent{\bf AC power flow model:} We use the SOCP relaxation of the AC power flow for radial distribution networks~\cite{low_exact_relaxation} in the same manner as the setpoint algorithm in~\cite{CDCp}.

	\noindent{\bf Nomenclature:} Table~\ref{varTable} introduces the nomenclature used throughout this paper.
	
	\begin{table}[!htbp]
		% \resizebox{\textwidth}{!}{%
		\vspace{-5mm}
		\caption{Variables and constants.}
		\vspace{-2mm}
		\label{varTable}
		\begin{tabular*}{0.5\textwidth}{r p{5.7cm}}
			\hline
			\textbf{Symbol} & \textbf{Description}  \\ \hline
			$t, T, k, i, j, N$  & time, timesteps in horizon, battery index, node indices, nodes in network \\
			$\mathcal{Z}, \mathcal{E}, E$ & set of controllable nodes, set of EVs, set of network edges \\
			$s, x$ & net complex power injection, real part of $s$ which is the upper or lower bound \\
			$w, W, y, W_{tol}$ & node voltage squared, voltage squared semidefinite matrix, line impedance, squared voltage limit \\
			$p^{\mathrm{target}}, p^{\mathrm{base}}, s^{\mathrm{forecast}}$ & target real power injection, baseline real power injection, baseline complex power injection \\
			$\lambda_v, \lambda_d, I_b$ & cost of voltage deviation, cost of baseline deviation, indicator of upper or lower bound \\
			$c, d, Q, p$ & battery charge power, discharge power, available capacity, local power forecast \\
			$\lambda_e, \lambda_b, \gamma_l, \gamma_c, \gamma_d$ & energy cost, battery operation cost, battery leakage, charge, and discharge efficiency \\ 
			\hline
		\end{tabular*} %
		\vspace{-5mm}
	\end{table} 
	
\section{Coordination scheme} \label{Architecture}
\vspace{-1mm}
	
	The coordination scheme consists of a day-ahead scheduling phase run by the GC, for example the grid operator, to determine power injection upper and lower bounds, and an operating phase run entirely by the LCs. Each LC optimizes its objective in real time with a high penalty for violating the power injection bounds sent by the grid operator. 
\vspace{-1mm}
\subsection{Day-ahead power bounds scheduler}
\vspace{-1mm}
First, we describe the day-ahead scheduling of the power injection bounds run by the GC.

	% Another way to submit targets adds the flexibility of the DERs to the baseline
	% Do not discuss any pros and cons of different ways to do the targets instead say our focus is on the use of power bounds to protect grid assets
	% can be a topic of further investigation in another paper
\noindent{\bf Determining upper and lower power targets}: The scheduler runs each day to determine the power bounds for each participating node in the network for the following day. The scheduler first specifies an upper and lower target value for the power injection at each participating node defined as $p^{\mathrm{target}}$. There are several ways to determine these targets. 

\noindent{1.} The targets are determined through a day-ahead market for the DER operators to bid on the amount of flexibility and capacity they would like to be given by the grid---higher flexibility means a larger gap between the upper and lower power targets and higher capacity means a larger upper target. Low flexibility operation would reduce the uncertainty of the grid operator, which would reduce the cost of balancing power in the system. Additionally, there is a limit to the amount of power each node under the same transformer can consume. Nodes that would like to consume more power would have to pay for a higher upper target. The auction will have time varying prices that balance the amount of power across the network so that each node could potentially find a time period to consume more power than what would have been previously supported by the network. %This is possible because we can dynamically reallocate additional power capacity to the nodes who are willing to pay for it. This enables the grid operator to support more loads such as EV charging without needing as many equipment upgrades. 

\noindent{2.} The targets are determined by adding the flexibility of the DERs, e.g. storage charging capacity, to a baseline estimate.

\noindent{3.} The targets are determined using historical data to forecast the day-ahead power consumption. This method can be performed by the grid operator using AMI data, but it may be exploitable by the DER aggregators who can deliberately increase their uncertainty to be awarded with more flexibility.

A deeper exploration of methods for selecting the upper and lower power injection targets is a topic for further investigation since the goal of this paper is to determine if day-ahead upper and lower bound scheduling can achieve similar performance to setpoint approaches with significantly less information. Toward this goal, we use the last method in which we add/subtract one standard deviation of forecast error to the forecast to obtain the upper/lower bounds for the experiments in this paper.
Finally, the upper and lower targets are compared to the two-way static transformer power ratings, and the tighter of the two are input to the bounds optimization. This ensures that the nodes are not assigned bounds that are beyond the ratings of their transformer.
	
	%\noindent{\bf Discussion of targets:} %For realistic radial distribution grids, the bus voltage magnitude is generally an increasing function of power injection at each connected bus. Therefore, bounding the power injection implicitly bounds the voltage magnitude. We present a brief and simple look at the relationships between power and voltage through the power flow equations with realistic distribution grid parameters in the appendix. Also, we show empirically that the power bounds implicitly bound the voltage in the network in Section~\ref{Results}.
	
	%Although calculating the bounds assuming the maximun or minimum power injections for each node occur simultaneously leads to a conservative bound, we find that it is not too conservative so as to obstruct the local device objectives and is shown in Section~\ref{Results}. This is opposed to the method of calculating bounds in~\cite{kyle} and~\cite{CDCp} where other nodes are assumed to be following their optimal trajectory, while we calculate the bounds for only a single node. In the previous work, we found that method of calculating bounds to be too loose and did not provide much value in many cases, while the worst case consideration is much safer while not obstructing local DER objectives.
	
	\noindent{\bf Network power bounds optimization:}
	To determine the upper and lower bounds on power injection for each participating node in the network, the scheduler solves two optimization problems. The objective functions for these problems combine terms to determine the upper or lower power injection bounds based on the targets and the sum of the squared voltage deviations over all buses proposed in~\cite{CDCp} to penalize bus voltage magnitudes that are out of the desired range. 
	 
	 More specifically, the scheduler solves the following optimization problem once to determine the lower bounds and again to determine the upper bound, each time solving for $x \in \mathcal{R}^{N\times T}$, which is the bound for all nodes over the time horizon.
	
	{\allowdisplaybreaks
	\begin{subequations}
		\vspace{-5mm}
		\begin{alignat}{2}
		\underset{x,w}{\min} \quad & \lambda_v \sum_{i=0}^{N} \sum_{t=0}^{T} ( \lbrack w_{it} - W_\mathrm{tol+} \rbrack_+ + \lbrack W_\mathrm{tol-} - w_{it} \rbrack_+ )^2 \label{v_penalty} \\
		& + \sum_{i=0}^{N} \sum_{t=0}^{T} (x_{it} - p_{it}^{\mathrm{target}})^2 \label{projection} \\
		& + \lambda_d \sum_{i=0}^{N} \sum_{t=0}^{T} \lbrack I_b (x_{it} - p_{it}^{\mathrm{base}}) \rbrack_+ \label{direction} \\
		%\nonumber \\
		\text{s. t.} \quad & s_{it} = s_{it}^{\mathrm{forecast}} \hspace{1em} \forall i \notin \mathcal{Z} \label{baseline} \\
		& x_{it} = \Re(s_{it}) \label{real_part} \\
		%& \boldsymbol{v} = \boldsymbol{A} \begin{bmatrix}\boldsymbol{x} \\ \boldsymbol{q} %\end{bmatrix} + \boldsymbol{b}. \label{linearPF}
		& s_{it} = \sum_{j:(i,j)\in E} (w_{ijt} - w_{iit})y_{ij}^{*} \label{GCopt7} \\
		& W\{i,j\}_t \succeq 0. \label{GCopt8}
		\end{alignat}
		\vspace{-3mm}
	\end{subequations}
	}
	
The term~\eqref{v_penalty} in the objective function is the voltage penalty and has a steep weight $\lambda_v$. It replaces the more commonly used hard voltage constraints in order to avoid excessive tightening of the bounds and solution infeasibility, see~\cite{CDCp}. The term~\eqref{projection} aims to find a feasible set of power injection points in the network that is close to the nodes targets without incurring excessive voltage penalty. The term~\eqref{direction} provides a penalty to discourage sensitive nodes in the network from having bounds that do not provide a minimum capacity given by $p^{\mathrm{base}}$ . The~\eqref{baseline} constraint assumes that the non participating nodes (nodes not in set $\mathcal{Z}$) are assigned their forecasted power consumption. The variable $I_b$ is equal to $1$ for the upper bound and $-1$ for the lower bound and determines the direction of the bound. 

Equations~\eqref{GCopt7} and~\eqref{GCopt8} are the convex relaxation for the power flow equations, where $W$ represents the rank $1$ semidefinite matrix of voltages (see~\cite{low_exact_relaxation} and~\cite{CDCp} for more details). Note that the forecasts themselves do not appear in the scheduler equation directly, but are part of the target power injections $p^{\mathrm{target}}$ calculated as previously described. DER constraints are deliberately not included in the optimization because we wish to find power injection bounds with respect to voltages regardless of resource capabilities, which are handled by the local controllers.

This method of calculating bounds compares to our previous work in the following ways. The method of calculating bounds in~\cite{kyle} and~\cite{CDCp} was for a single node at a time and assumed all other nodes are following their optimal trajectory. These bounds supplemented the setpoint tracking to provide flexibility; however, we found that they are often loose. Our scheduler considers a scenario in which we assume the maximum or minimum power injections for all nodes occur simultaneously when determining the targets. Although the use of these targets results in conservative bounds, we find that they are not so conservative as to adversely affect the local DER objectives for this case as shown in Section~\ref{Results}.
	
	%Note that we make the assumption that the upper or lower bound is calculated based on all of the nodes simutaneously providing excess generation or consumption. Although this is a conservative assumption that leads to a conservative bound, we find that the bound is not too conservative as to restrict the local objectives of the devices and is shown in Section~\ref{Results}. In~\cite{CDCp}, we found the assumption that all nodes except one operating optimally was too loose and did not provide a tight enough bound to be meaningful in many instances; therefore, in this paper we explore the assumption that all nodes are deviating simutaneously.
	
	\vspace{-1mm}
	\subsection{Local controllers}
	\vspace{-1mm}
	During the operating phase, each LC operates in a rolling horizon fashion. The objective is to minimize local costs while remaining within the bounds provided by the day-ahead scheduler. The output is the storage charging profile to be executed by the storage unit, if any, and the charging profile of any EVs under the LC's control. The formulation is as follows:
	
	{\allowdisplaybreaks
		\begin{subequations} \label{Localopt}
			\vspace{-5mm}
			\begin{alignat}{2}
			\underset{c,d,Q}{\min} \quad & \sum_{t=0}^{T} \lambda_e \big[ p_t + \sum_{k=0}^{K}(c_t^{(k)} - d_t^{(k)}) \big]_+ \label{energy_cost} \\
			%\nonumber \\
			%\nonumber \\
			& + \lambda_v \sum_{t=0}^{T} ( \lbrack p_{t} - x_+ \rbrack_+ + \lbrack x_- - p_{t} \rbrack_+ )^2 \label{bounds_penalty} \\
			& + \lambda_b \sum_{t=0}^{T} \sum_{k=0}^{K} (c_t^{(k)} + d_t^{(k)}) \label{battery_penalty} \\
			\text{s. t.} \quad & 0 \le c \le c_{\mathrm{max}} \label{cmax} \\
			& 0 \le d \le d_{\mathrm{max}} \label{dmax} \\
			& Q_t = \gamma_l Q_{t-1} + \gamma_c c_t - \gamma_d d_t \label{charge} \\
			& Q_{\mathrm{min}} \le Q \le Q_{\mathrm{max}} \label{qmax} \\
			& Q_{t=\mathrm{end}}^{(k)} = Q_{\mathrm{final}}^{(k)} \hspace{1em} \forall k \in \mathcal{E}. \label{ev_charge}
			\end{alignat}
			\vspace{-4mm}
	\end{subequations} 
}
	
	The term~\eqref{energy_cost} in the objective function is the cost of energy without any net metering incentives (we assume that there is no profit for selling energy back to the grid). This term, however, can be customized based on the DER objectives. The term~\eqref{bounds_penalty} is a quadratic penalty for the power consumed above or below the bounds, and its minimization enables the LCs to respect the bounds sent from the GC. The term~\eqref{battery_penalty} discourages excessive wear on the battery with cost $\lambda_b$. Since the other terms in the objective function consider only the difference between $c$ and $d$, charging and discharging cannot simultaneously occur as such simultaneity would increase the cost. The battery variables are indexed by $k$ for each EV or stationary battery under an LC. The variable $p_t$ represents the local power forecast at time $t$. The forecast can be improved by the addition of forecast scenarios as in~\cite{kyle}, but we decided not to include scenarios here for simplicity of notation. The battery operating constraints~\eqref{cmax}~\eqref{dmax}~\eqref{qmax} are for the capacity limits defined as $c_{\mathrm{max}}, d_{\mathrm{max}}, Q_{\mathrm{max}}$, respectively, and constraint~\eqref{charge} defines the battery dynamics. Equation~\eqref{ev_charge} requires the charge capacity of each EV (set of all EVs is $\mathcal{E}$) to reach the final value by the end of the charging period to ensure quality of service. The battery operation constraints are vectors with elements representing EVs and stationary battery under an LC. We do not consider vehicle to grid interaction, hence $d_{max} = 0$ for all EV batteries. We use two battery charging parameters $c$ and $d$ to enable more accurate modeling of efficiency. The objective is minimized when either $c$ or $d$ is 0 since only the difference between them is considered while their sum is penalized in~\eqref{battery_penalty}. %Other controllable devices such as thermostats, PV, or generators, can be included, and different objectives can be run for each LC without any needed changes at the global level.
	
	%\vspace{-2mm}
	\vspace{-1mm}
	\subsection{Benchmarks}
	\vspace{-1mm}
	We compare the performance of our scheme to three different benchmarks. The first is with LCs operating their DERs independently without any bounds signals. The second is with the LCs operating with static bounds determined by transformer power capacity ratings. The only difference between these benchmarks and our method is the value of the bounds in the local objective. The final benchmark is the global setpoint tracking scheme specified in~\cite{CDCp}.
	
	\vspace{-3mm}
	\section{Simulation results} \label{Results}
	\vspace{-1mm}
	
	We perform simulations using the IEEE 123-bus network~\cite{IEEE_feeders} including unidirectional voltage regulators and capacitor banks over two summer months of power injection data provided by Pecan Street using the MATPOWER~\cite{zimmerman2011matpower} power flow simulator. The solar data is from NREL~\cite{NRELsolar}, and the forecaster is the ARIMA model used in~\cite{CDCp}. The EVs are sampled from probability densities as explained in Section~\ref{models}. We use hourly time resolution in both the GC and LCs to match existing day-ahead markets, and the total optimization horizon is 2 days (48 points). DER owners in this network pay a time of use energy price with a peak price of 46.3 cents per KWh from 4pm to 9pm and to 20.5 cents per KWh for the rest of the day. %The upper and lower target power injections are determined by forecasting the day-ahead power injection for each participating node and adding/subtracting one standard deviation of forecast error to obtain the upper bound/lower bound. %Including large EV charging spikes in the demand has made the forecaster less accurate, so one standard deviation represents a large amount of flexibility during hours with large amounts of EV charging.
	
	The metrics we choose to evaluate the results are (i) arbitrage profit, (ii) the voltage deviation metric defined in~\cite{CDCp}, and (iii) a transformer capacity deviation metric. Arbitrage profit is the money saved by the batteries buying electricity during the low price period and offsetting demand during the peak price period. The voltage deviation metric, which is the sum of squared voltage deviation above the nominal range, is in the objective funcion of the GC while the transformer capacity deviation metric, which is the sum of squared power deviation beyond the rating, is in the objective function of the LC. The computation time for a single LC on a personal computer is approximately 0.15 seconds, which demonstrates its feasibility for use in real time control. Next, we evaluate the performance metrics through simulation of different DER scenarios.
	
	\noindent{\bf Network without storage:}
	We begin by examining the voltage deviation and transformer capacity deviation caused by the addition of DERs with no storage. Table~\ref{baseTable} shows that before the addition of any DERs, the network operates smoothly without voltage or transformer violations, but adding solar increases voltage deviations due to back-feeding during peak sunlight hours. Adding EV charging to the network also increases transformer capacity deviations especially due to the high power consumption requirements fast charging. When both types of DERs are included, they can help balance each other somewhat; however, the performance can be greatly improved through the coordination of distributed storage as will be shown below.
	
%	\begin{table}[!t]
%		\centering
%		\vspace{5mm}
%		\includegraphics[width=3in]{baselineTable.png}
%		\caption{Voltage deviation and transformer capacity deviation metrics with the addition of DERs with no storage.}
%		\label{baseTable}
%		\vspace{-9mm}
%	\end{table}

\begin{table}[!t]
	\caption{Voltage deviation and transformer capacity violation metrics with the addition of DERs with no storage.}
	\vspace{-2mm}
	\label{baseTable}
	\begin{tabular}{|l|r|r|}
		\hline
		Scenario            & \multicolumn{1}{l|}{Voltage Deviations} & \multicolumn{1}{l|}{Transformer Violations} \\ \hline
		No DERs             & 0                                       & 0                                           \\ \hline
		Solar = 50\%        & 3.649                                   & 0                                           \\ \hline
		EV = 50\%           & 0                                       & 4.48 E6                                     \\ \hline
		Solar and EV = 50\% & 1.608                                   & 3.385 E6                                    \\ \hline
	\end{tabular}
\vspace{-7mm}
\end{table}
	
	\noindent{\bf Aggregate performance metric vs. Quantity of storage:}
	Here we compare the performance of the setpoint tracking scheme and the two benchmarks to our proposed method by simulating the control for each case and evaluating the voltage deviation and transformer capacity deviation metrics at varying penetrations of storage and 50\% penetration of solar and EVs. The storage is co-located with the solar generation and the capacity is proportional to the power demanded by the node connected to it. The penetration is defined as the percentage of the total network daily energy that can be stored. 
	
\begin{figure}[!t]
		\centering
		\includegraphics[width=3in]{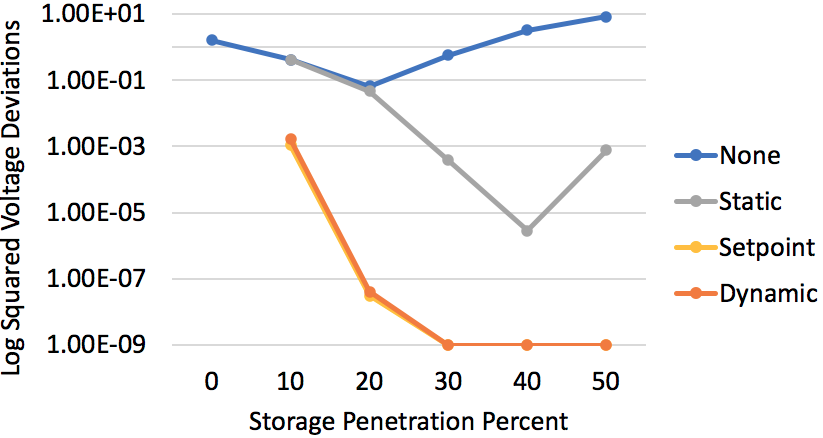}
		\vspace{-2mm}
		\caption{Sum voltage deviations versus storage penetration on a log scale. Dynamic bounds achieves similar performance to setpoint tracking and significant improvement over the 2 benchmarks. The two benchmarks' performance becomes worse with large storage capacity as the storage itself begins to cause voltage violations.}
		\label{logVoltage_stor}
		\vspace{-5mm}
	\end{figure}

	Figure~\ref{logVoltage_stor} plots the voltage deviation metric. Note that the voltage deviations are approximately the same for the no bounds and the static bounds cases except when storage capacity is large. They are significantly lower, however, for our proposed dynamic bounds. The metric for the two benchmarks becomes worse with high storage as the storage itself begins to cause voltage violations. We can see that the dynamic bounds can use storage better, enabling significant performance increase with less storage than with the static bounds. This demonstrates that the dynamic bounds act as implicit bounds on the bus voltage magnitudes and that transformer capacity ratings alone is not sufficient to keep the voltages in the grid within acceptable bounds. The performance of the dynamic bounds matches that of the setpoint tracking scheme.
	
\begin{figure}[!t]
		\centering
		\includegraphics[width=3in]{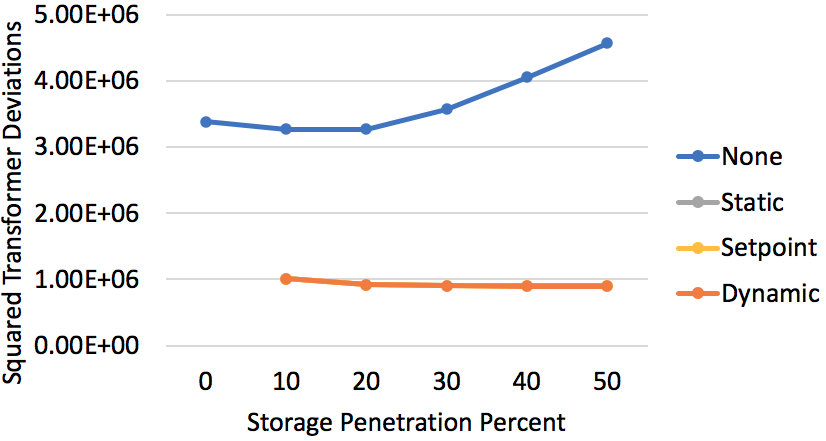}
		\vspace{-2mm}
		\caption{Transformer capacity deviations vs. storage penetration. The inclusion of bounds in the LC greatly improves performance, although it is the same for either static bounds, dynamic bounds, or setpoint tracking. The violations are due to the fast charging stations and lack of sufficient storage capacity for the highest power peaks, thus, transformer upgrades are unavoidable.}
		\label{tVio_stor}
		\vspace{-5mm}
	\end{figure}

	Figure~\ref{tVio_stor} plots the transformer deviation metric. It shows that adding transformer capacity bounds to the LCs can offer a significant improvement over the no bounds case. Note that the plots for all cases except for the no bounds benchmark coincide because the IEEE 123-bus network is a residential network in which all transformers are secondary transformers, thus, the power flowing through them only serves their loads and not other nodes. The majority of deviations are due to the fast charging stations and lack of sufficient storage capacity to protect from the highest power peaks, and not a limitation of the LC algorithm.
	
\begin{figure}[!t]
		\centering
		\includegraphics[width=3in]{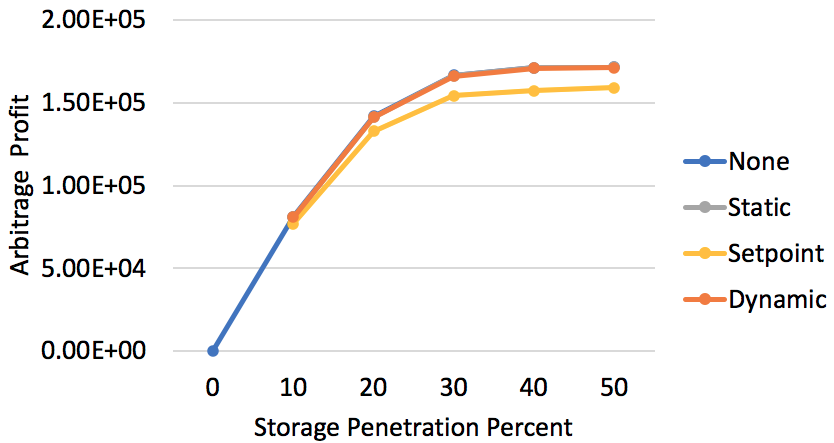}
		\vspace{-2mm}
		\caption{Arbitrage profit vs. storage penetration. The dynamic bounds has similar arbitrage profit as both benchmarks, demonstrating that the bounds do not hinder the LC objectives. The setpoint tracking scheme achieves lower arbitrage profit because the setpoint is calculated based on delayed global forecasts and can be inaccurate, while in the other cases, the LC uses local data to compute its power injection trajectory.}
		\label{arb_stor}
		\vspace{-5mm}
	\end{figure}

	Figure~\ref{arb_stor} plots the arbitrage profit for the storage units that perform cost minimization. Note that the two local bounds test cases have nearly the same arbitrage profit as the no bounds case, which shows that the use of bounds in the LC does not hinder the storage units ability to perform arbitrage. Although the dynamic bounds are conservative due to the assumption that all nodes experience their peak power injection simultaneously, they do not adversely affect the local objectives. The setpoint tracking performance is slightly worse because the setpoint is calculated using global forecasts, which can be inaccurate, whereas the other cases use accurate local measurements to calculate their objective within bounds~\cite{kyle}. Through these simulations, we can see that the dynamic bounds scheme is able to match the performance of the best case for all three metrics, making it a better scheme than any single method overall.
	
\begin{figure}[!t]
		\centering
		\includegraphics[width=3in]{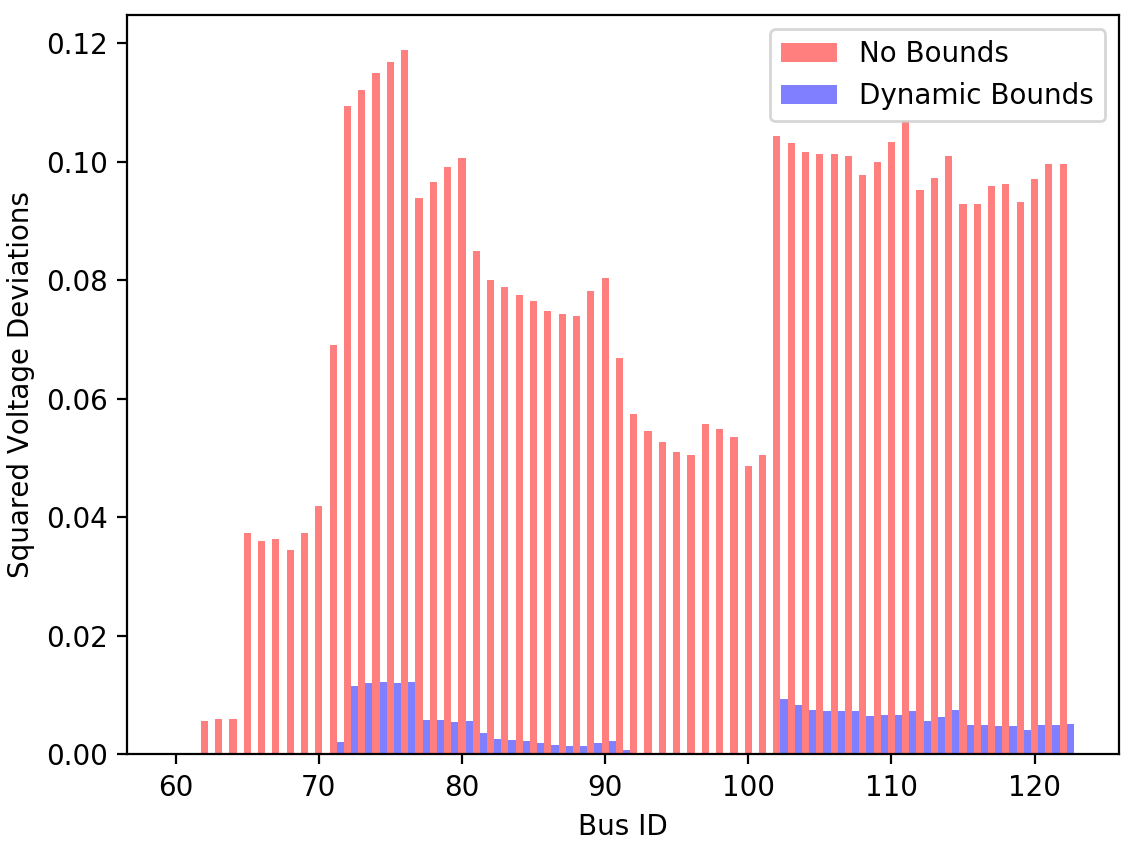}
		\vspace{-2mm}
		\caption{Voltage deviations at each bus for the different cases. They are approximately equal for the no bounds and static bounds cases, but are significantly lower for our proposed dynamic bounds and setpoint tracking. The first 60 buses contain no voltage deviations as they are close to the substation.}
		\label{bus_voltage}
		\vspace{-5mm}
	\end{figure}

	\noindent{\bf Performance metrics per bus:}
	Here we take a look at the performance metrics for each bus individually. Figure~\ref{bus_voltage} plots the voltage deviation metric per bus with 10\% penetration of storage. The voltage deviation metric per bus is approximately the same for the no bounds and static bounds cases, but are significantly lower for our proposed dynamic bounds and the setpoint tracking scheme. As expected, the unidirectional voltage regulator located at bus ID 72 causes more severe voltage deviations at neighboring buses due to the voltage boost it provides during solar back-feeding.
	
\begin{figure}[!t]
		\centering
		\includegraphics[width=3in]{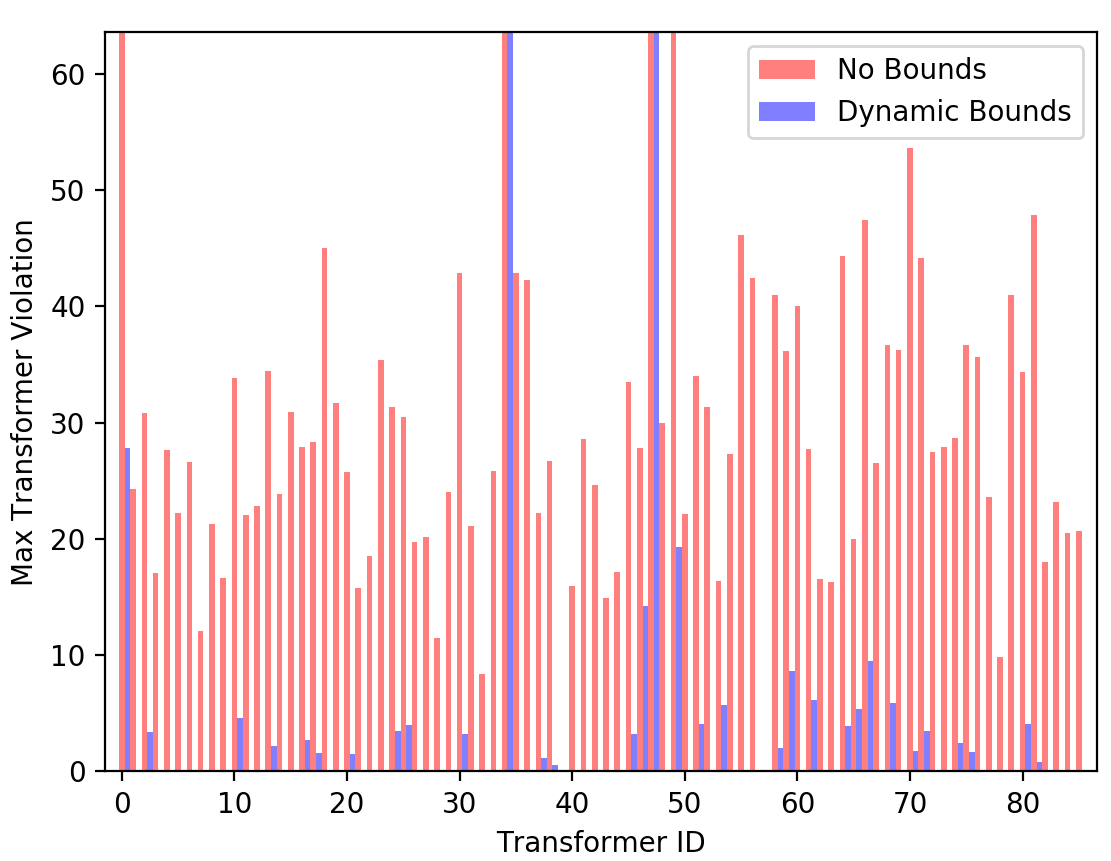}
		\vspace{-2mm}
		\caption{Transformer capacity violations at each bus for the different cases. There are large capacity violations across the network with no power bounds, but dramatically less violations with bounds or setpoints. The four peaks that are cut off correspond to the fast charging stations, whose capacity violations exceed the transformer capacity by over 150 kW.}
		\label{bus_tVio}
		\vspace{-6mm}
	\end{figure}

	Figure~\ref{bus_tVio} shows the maximum transformer capacity violation (in kW) at each secondary transformer in the network. The violations are significantly worse when there is no control over the maximum power allowed to flow through the transformer as demonstrated by the no bounds case; however, the dynamic bounds, static bounds, and setpoint tracking cases greatly reduce the maximum capacity violation. We can see that using bounds can avoid upgrading many transformers that would otherwise need to be replaced due to the addition of DERs. Only 12 transformers incur power violations greater than a single EV charger (4 kW) with bounds, while nearly all 86 transformers have large violations without any bounds. The four highest peaks are the fast charging stations.

	\vspace{-2mm}
	\section{Conclusion}
	\vspace{-1mm}
	%Here we discuss a few interesting aspects of the algorithms that may arise as common questions. 
	%\begin{itemize}
	%\item The experiments here are performed on a single phase system, but we see no reason why it would not work on an unbalanced 3-phase system with the 3-phase power flow equations and bounds on each phase. The evaluation of such a system is the subject of future work.
	%\item If the network line parameters are unknown, a linear or input convex model relating the power injection to the bus voltages can be learned from AMI data and used to replace the power flow constraints~\cite{LinearPF_LSQ, LinearPF_Jiafan_older, LinearPF_Jiafan}.
	%\item It is possible to add other costs to the GC objective for distribution network reliability such as costs associated with violating line ampacities.
	%\item It is possible to add ancillary service following to the objectives of the DERs in the LC. The scheduling of ancillary service signals can be aided by the bounds calculated by the GC. The formulation of such an algorithm is the subject of future work.
	%\end{itemize}
	We have presented a DER coordination method that involves a day-ahead global scheduling of power injection bounds to reduce voltage deviations and transformer capacity violations in a distribution grid, with only smart meter data and no required knowledge of the DER owners' data or objectives. The local controllers operate their DERs autonomously with a penalty for deviating from the globaly scheduled bounds. We have demonstrated via simulations on the IEEE 123 bus network that the addition of distributed solar can cause voltage deviations due to back-feeding, and the addition of EV charging can add transformer capacity violations, and the coordination of storage and EV charging through the use of our dynamic bounds scheme can greatly mitigate these violations while maintaining unimpeded performance of the DERs' objective of arbitrage profit. Our dynamic bounds scheme matches the best performance of any of the benchmark cases for the three metrics, making it a better scheme than any single method presented overall.
	
	%Additionally, the global scheduler does not require any knowledge of the DERs in the network and only needs to communicate daily schedules to the nodes, which can enable cooperation with day-ahead markets. 
	The method we described allows for the addition of other controllable devices such as thermostats, PVs, or generators, and different objectives can be run for each LC without any needed changes at the global level. It is possible to add other costs to the GC objective for distribution network reliability such as costs associated with violating line ampacities. Finally, potentially over conservative bounds can be loosened by performing a Monte Carlo simulation over a distribution of target forecasts instead of using only the maximum (or minimum) power injections over all the nodes.
	
	\vspace{-2mm}

	\bibliographystyle{IEEEtran}
	\bibliography{references_NBC.bib}

\end{document}